# GaN / VO$_2$ heteroepitaxial p-n junctions: Band offset and minority carrier dynamics

You Zhou and Shriram Ramanathan

*Harvard School of Engineering and Applied Sciences, Harvard University, Cambridge, Massachusetts 02138, USA*

**Abstract**

We report on experimental realization of p-n heterojunctions based on p-type GaN, and an n-type correlated oxide, VO$_2$. The band offsets are evaluated by current-voltage and capacitance voltage measurements at various temperatures. A band diagram based on the conventional band bending picture is proposed to explain the evolution of the apparent barier height from electrical measurements and it suggests that the work function of VO$_2$ decreases by ~0.2 eV when it goes through the insulator to metal transtion, in qualitative agreement with Kelvin force microscopy measurements reported in literature. The frequency-dependent capacitance measurements allows us to differentiate the miniority carrier effect from the interface states and series resistance contributions, and estimate the minority carrier lifetime in insulating phase of VO$_2$ to be of the order of few microseconds. The nitride-oxide based p-n heterojunctions provide a new dimension to study correlated-electron systems and could be of relevance to emerging electronic devices that exploit collective phenomena.

# I INTRODUCTION

Correlated oxides exhibiting metal-insulator transitions due to the interplay of electronic and lattice degrees of freedom are being extensively investigated in bulk and thin film forms in the condensed matter sciences community. P-n junctions are among the building blocks of semiconductor and energy devices, such as but not limited to diode transistors, metal-oxide-semiconductor field-effect transistors (MOSFETs), light emitting diodes (LEDs) and solar cells. Fabricating p-n junctions with correlated oxides allows one to dynamically control the carrier density in the space charge region within the correlated oxides and study how the correlated physics evolves with the carrier density (a schematic is shown in figure 1).[1, 2] These devices also provide an elegant approach to look into some physical properties such as minority carrier charge/spin dynamics and electron-photon interactions in such materials,[3-6] and to estimate band diagrams under electric fields.[7-9] The tunable conductance states, inherent memory make such materials and their junctions rather interesting candidates for new classes of reconfigurable electronics, learning circuits and in a broader setting as adaptive devices.

Vanadium dioxide ($VO_2$) (monoclinic and rutile structure in the insulating and metallic phase, respectively) is of interest for electronic and photonic devices since it shows a metal-insulator transition with sharp change in the electric and optical properties which can be triggered by temperature, voltage and optical fields at room-temperature, while its transition mechanism is still under intense study. Integrating vanadium oxides into hetero p-n junctions may not only contribute to the understanding of the metal-insulator transition in the VO$_2$, but may also give rise to electronic/optical devices with new functionalities. However, it is non-trivial to fabricate VO$_2$ based junctions with rectifying transport properties because of its crystal symmetry, comparatively high work function[10-13] and high carrier density even in its insulating phase[14, 15]. To the best of our knowledge, only isostructural Nb-doped TiO$_2$/VO$_2$ p-n junctions have been demonstrated to study photo-carrier injection junctions until present.[16, 17] In the current study, we investigate the electronic properties of the junction formed between n-type VO$_2$ and a p-type conventional semiconductor, GaN. GaN has important applications for LEDs, lasers and high-power electronics and could serve as a substrate for the epitaxial growth of VO$_2$. Another advantage provided by GaN is that its Fermi level is tunable across a wide bandgap through doping, providing large enough band bending and space charge region in the GaN/VO$_2$ junction.[18, 19] Rectifying current-voltage characteristics of such a heterojunction are demonstrated in both the insulating and metallic phase of VO$_2$. We investigate the band offsets at the junction by electrical measurements and explain the observations within the framework of a band bending picture by considering the *3d* bands of VO$_2$. The minority carrier lifetime in VO$_2$ is extrapolated from the frequency-dependent capacitance measurements in different DC bias regime.

## II    EXPERIMENTS

Commercially available Mg-doped *p*-type GaN on c-plane sapphire grown by MOCVD was used in this study (from University Wafer). The carrier concentration is ~ 5 × 10$^{17}$ cm$^{-3}$ from Hall measurements. Prior to the VO$_2$ deposition, the GaN substrates were degreased consequently in acetone and isopropanol for 5 mins. Then the substrates were treated with boiling aqua regia ($HNO_3$:HCl = 1:3) to remove surface oxide.[20] Afterwards, vanadium dioxide films of various thicknesses were grown on GaN by magnetron sputtering from a $V_2O_5$ target in an Ar/$O_2$ gas mixture of a total pressure of 5 mTorr at 500 ºC. The $O_2$ partial pressure was kept at 0.013 mTorr during the deposition. We note that the formation enthalpy is lower for $VO_2$ than $Ga_2O_3$ above room temperature. X-ray diffraction studies and pole figure analysis have confirmed that the $VO_2$/GaN heterostructure has an epitaxial relation of (out-of-plane) $VO_2(010)_R$//GaN(0001) and (in-plane) $VO_2[100]$//GaN[$10\bar{1}0$] (The subscript R denotes the rutile phase of $VO_2$).[21] The as-grown $VO_2$ films were patterned into squares of various sizes and square shaped Ti(20 nm)/Au(200 nm) ohmic contacts to $VO_2$ were deposited by sputtering with standard photolithography techniques. In order to make low resistance ohmic contact with p-type GaN, the surface of GaN was treated with boiling aqua regia for 10 mins prior to the deposition of Pd(10 nm)/Au(30 nm) contact by electron beam evaporation and lift-off.[20] The fabricated Pd/Au contact to p-type GaN shows ohmic contact behavior in the temperature range of the present study. Hall measurements were done with MMR Hall Measurement System in Van der Pauw geometry. Temperature-dependent current-voltage and capacitance-voltage characteristics were measured by Keithley 2635A sourcemeter and Agilent 4980A Precision LCR Meter with samples placed on a heating chunk in dark. The electrical measurements were performed in dark environment.

# III      RESULTS AND DISCUSSON

## A. Band diagram of $VO_2$/GaN heterojunction

Before discussing the band diagram of the $VO_2$/GaN heterojunction, it is essential to provide a description of the electronic band structure of both materials. GaN is a conventional semiconductor with a band gap of 3.4 eV at room temperature[22] and the change in band gap within the temperature studied here is less than 0.1 eV.[23] Its band structure could be described by weakly-interacting electron band theory. On the other hand, $VO_2$ is a transition metal oxide with a $d^1$ electron configuration. The electrons in narrow $d$ band could be strongly correlated, making it challenging to calculate the band structure of $VO_2$ (especially for the insulating phase) *ab initio*. The following section gives a brief description of its band structure based on experiments and first principle simulations reported in literature.

Above the transition temperature, $VO_2$ is metallic and has a rutile crystal structure, in which the vanadium atoms are located at the center of the oxygen octahedra. According to crystal-field theory, the 3d states are split into lower sixfold degenerate $t_{2g}$ bands and higher fourfold degenerate $e_g$ bands (including spin degeneracy), whereas the filled oxygen *2p* band lies well below (~ 2.5 eV below) the $d$ band.[24, 25] Because there is only one $d$ electron per vanadium, there is negligible contribution to electrical conductivity from $e_g$ bands and thus only $t_{2g}$ bands will be considered in the following discussion. In the rutile structure, each $VO_6$ octahedron shares edges with its two neighboring octahedra along the rutile *c* axis and shares vertices with the neighbors in other directions, forming chains parallel to the $c$ axis. Such specific stacking of oxygen octahedra will lead to a further splitting of the $t_{2g}$ bands into a lower twofold degenerate $d_{//}$ band and an upper fourfold degenerate $\pi$ band, which is shown in Fig 2(a).[26] The $d_{//}$ band is derived from the local vanadium $3d$ orbital whose electron density is directed along the c-axis and has quasi-one dimensional dispersion parallel to c axis. On the contrary, the $\pi$ band is originated from the other two 3d orbitals in the $t_{2g}$ bands with a rather isotropic dispersing relation. The $d_{//}$ and $\pi$ bands have a total bandwidth of ~ 2 eV and the Fermi level lies ~0.6 eV above the conduction band minimum in the $d$ band.[26-29] Therefore, a schematic band structure of metallic $VO_2$ could be constructed as in Fig. 2(a) based on the above discussion.

The electronic band structure of the insulating phase is closely related to that of the metallic phase. When cooled below the transition temperature, $VO_2$ transforms into a monoclinic structure, which is characterized by the dimerization of vanadium atoms along the rutile $c$ axis and the tilting of these vanadium dimers perpendicular to $c$ axis. In the Goodenough picture based on weakly-interacting electron theory, the dimerization leads to the splitting of the quasi-1D $d_{//}$ band as in Peierls transition and the tilting results in the upshift of $\pi$ bands.[30] This opens a band gap in the monoclinic phase and $VO_2$ becomes a semiconductor. In the Mott-Hubbard model proposed by Zylbersztejn and Mott,[31] on the contrary, the formation of the band gap is due to strong electron correlation within the narrow $d_{//}$ band that splits itself into an upper and a lower Hubbard band, $d_{//}^*$ and $d_{//}$. In this strong correlation picture, when correlated $VO_2$ goes above the transition temperature, the $\pi$ bands are energetically lowered by the structural transition and become filled with electrons. The electrons in the filled $\pi$ bands can therefore screen the electron correlation in $d_{//}$ band, merging the two $d_{//}$ Hubbard bands into one and resulting in an insulator-to-metal transition. Despite the difference of these two scenarios in explaining the metal-insulator transition mechanism, the band structure of the insulating phase (density of states) that is of interest for the current study could generally be presented as in Fig. 2(b). The $d_{//}$ and $d_{//}^*$ bands correspond to bonding and anti-bonding bands in the Peierls transition picture, or the lower and upper Hubbard bands in the Mott transition picture. The band gap of the insulating phase is about 0.6 eV determined experimentally.[24, 27] The filled $d_{//}$ band is centered about 1eV below the Fermi energy with a band width of 1 eV.[27, 32-34] The total bandwidth of the conduction band ($d_{//}^*$ and $\pi$ band) is approximately 2 eV as estimated by first principle calculations.[29, 33] The energy separation between the O 2p band and $d_{//}$ band in the insulating phase is slightly reduced compared to the metallic phase, but still much larger than the ambient thermal energy $kT$.[35] The schematic band structure of the insulating $VO_2$ as described is shown in Fig. 2(b).

To construct the band diagram of the $VO_2$/GaN heterojunction, we assume an ideal case where there are no interface dipoles caused by surface atom arrangement (which could become important due to the ionic bonding nature of both $VO_2$ and GaN) or localized interfacial states (which could also be present because of the lattice misfit between $VO_2$ and GaN). In other words, the Anderson model for heterostructures (or Schottky model for metal-semiconductor contact) was employed for the following discussion (the electron affinity rule).[36-38] In such an ideal case, for the insulating phase of $VO_2$, the interface barrier height $\Delta E_C$ and $\Delta E_V$ are determined by the electron affinity $\chi$ of both materials, while the built-in potential $V_{bi}$ is related to their work function. For metallic $VO_2$/GaN interface, the barrier height $\Phi_b$ is given by $\Phi_b = \chi(GaN) + E_g(GaN) - \Phi_M(VO_2)$, and the band bending is related to the work function of p-type GaN. The reported electron affinity of GaN varies from 3.1 eV to 4.1 eV depending on different polar surface and characterization techniques.[39-42] The exact value of electron affinity turns out not to modify the band diagrams of the $VO_2$/GaN heterojunction as will be shown and therefore the often reported value of ~3.4 eV is assumed in the present study.[41, 42] The Fermi level in Mg-doped GaN with similar dopant concentration is ~ 0.1 eV above the valence band.[43] For $VO_2$, the evolution of work function across the phase transition has been reported recently. Therefore we can estimate the parameters such as electron affinity of $VO_2$ using the band structures from Fig. 2(a) and (b). However, previous studies have obtained subtly different results on the exact value and temperature dependence of the work function of $VO_2$.[10-13] For example, Ko et al. and Martens et al. have found that the work function evolves from 5.2 eV at room temperature to 5.3 eV at 100 ºC by Kelvin force microscopy and electrical characterization of $VO_2$-metal contact resistivity, respectively.[10, 11] On the other hand, Sohn et al. have shown that the work function drops from 4.88 to 4.70 eV when going through the insulator-to-metal transition.[12, 13] The differences in literature could be related to surface stoichiometry of VO$_2$ when exposed to air/humidity or in contact to another material and the measurement technique itself. The present study could in fact provide hints on the work function evolution of VO$_2$. Fig. 3(a) and (b) show the band diagrams of the VO$_2$/p-GaN junction for the metallic and insulating phase of VO$_2$, respectively. Notice that the O 2p band in both states are not drawn in Fig. 3, because it lies far below the Fermi level and would not contribute to conduction. As shown in Fig. 3(a), there is almost no band bending in metallic VO$_2$ because of its large carrier density (~ 10$^{23}$ cm$^{-3}$), leading to a Schottky barrier between the metallic VO$_2$ and p-GaN. The band bending in GaN is estimated to be ~1.3 - 2.1 eV and the Schottky barrier height is ~1.4-2.2 eV. Being different from a conventional metal, the metallic phase of VO$_2$ has a rather narrow conduction band (~ 2 eV) with low carrier mobility. This is highlighted in Fig. 3(a) where the bottom of the VO$_2$ *d* band is located near the Fermi surface and the density of states is zero below the *d* band. For the insulating phase, if we assume an ideal case of abrupt p-n junction with depletion approximation, the band bending in GaN will be ~1-2 eV, while the band bending in VO$_2$ is 0.1 ~ 0.2 eV. The built-in potential is ~1-2 eV and it drops mostly on the GaN side because of the comparatively large dielectric constant and carrier density of VO$_2$ ($\varepsilon_r$ ~ 30 - 40,[44, 45] electron density ~ 2 × 10$^{18}$ cm$^{-3}$ as measured by Hall effect in agreement with previous studies[14, 15]) with respect to GaN ($\varepsilon_r$ = 9.5,[46] hole density ~ 5 × 10$^{17}$ cm$^{-3}$ ). The band discontinuity at the interface is estimated by the electron affinity rule. The band diagram of the insulating VO$_2$/GaN heterojunction in Fig. 3(b) is similar to that of a type I straddling gap heterojunction, except for the narrow conduction and valence band of $VO_2$. In an ideal case, the injection efficiency of the p-n junction, defined as the ratio of the hole injection current into $VO_2$ to the electron injection rate of GaN, $i_n/i_p$, could be estimated to be extremely small because of the large difference between the band gaps of GaN and $VO_2$ ($\frac{i_{n1}}{i_{p2}} = \frac{D_{n1}}{D_{p2}} \frac{L_{p2}}{L_{n1}} \frac{N_2}{P_1} \left( \frac{m_{p1}m_{n1}}{m_{p2}m_{n2}} \right)^{\frac{1}{2}} \exp\left( \frac{E_{g2} - E_{g1}}{kT} \right)$, here 1 denotes GaN and 2 denotes $VO_2$. The injection efficiency is dominated by the band gap difference). Therefore the conduction of such a junction at forward bias would be dominated by hole injection from p-type GaN to $VO_2$. Similarly, the forward current across the junction above the phase transition temperature will also be primarily hole injection into $VO_2$.

### B. Current-voltage characteristics of the GaN-VO$_2$ junction

Fig. 4(a) shows the resistance of a ~30 nm thick $VO_2$ layer on p-type GaN as a function of temperature, which shows metal-insulator transition with a change in resistivity of about three orders of magnitude. The resistance was measured in-plane as schematically shown in the inset of Fig. 4(a). The current-voltage curves (from -0.1 to 0.1 V, not shown) are linear at all temperatures and the resistance values are calculated from linear fits. We note that the leakage current through the p-type GaN substrates would be negligible because of the rectifying behavior of the $VO_2$/GaN junctions as will be shown later. The transition temperature could be determined from the Gaussian fit of the dln$R$/d$T$ curves to be 75 °C and 68 °C for the heating and cooling branches, respectively as shown in Fig. 4(b). The width of the Gaussian peak, reflecting the temperature hysteresis width of the transition, is 17 ºC and 15 ºC for each branch. As previous studies point out, the transition temperature of VO$_2$ epitaxial layer could be tuned by the stress from the lattice mismatch with the substrate.[47] However, as the critical thickness of coherent growth for VO$_2$ is usually less than ~10 nm, most of the VO$_2$ layer is relaxed in the present study.[21, 48] In fact, VO$_2$ thin films with various thickness (16 nm to 67 nm) grown on GaN exhibit similar resistance-temperature characteristics, suggesting that the lattice is relaxed.

The typical current-voltage curves of the fabricated GaN/VO$_2$ heterojunction below and above the MIT transition temperature are shown in Fig. 5(a). The schematic of the fabricated devices is shown in the inset of Fig. 5(a). The electrodes are separated by ~ 500 μm in the fabricated devices. The *I-V* curves of all fabricated devices show rectifying behaviors in the temperature range studied. To eliminate possibility of Schottky contact formed at the Pd/GaN interface, the current transport properties through the Au/Pd contacts is studied as schematically shown in the inset of Fig. 5(a). As evidenced in Fig 5(b), the current voltage characteristics across two Au/Pd contacts (measured after the VO$_2$ deposition) are linear over the temperatures studied, indicating ohmic contact between Pd and p-type GaN. The current density in Fig. 5(b) is limited by both Pd/GaN contact resistance and bulk GaN resistance considering the large separation between the two Pd contacts (separated by ~500 μm). Similarly, the Ti/Au contact to VO$_2$ does not exhibit any non-linearity in its current-voltage curve, either. The above measurement verifies that the rectifying behavior is solely from the VO$_2$/GaN junction.

We note that the *I-V* curves of the p-n junction are not exponential but almost linear at large forward bias. This is primarily due to the comparatively large series resistance $R_S$ from the bulk GaN and Pd/GaN ohmic contact. As evidenced in Fig. 5(b), the current density across the Pd and bulk GaN contact at 5V is comparable to what is observed in Fig. 5(a). On the other hand, we can estimate the areal resistance of insulating $VO_2$ (25 nm thick, resistivity 10 Ω cm) to be $3\times10^{-5} \Omega/cm^2$, making it less likely to become the current limiting factor. Assuming that the $GaN/VO_2$ junction resistance is smaller than the series resistance $R_S$ at large forward bias, we could extrapolate $R_S$ from the linear fitting of the *I-V* curves at these bias and correct the voltage as $V' = V - IR_S$. The correction of series resistance $R_S$ makes the *I-V* curves sharper at forward bias and it is important for the analysis of ideality factor as will be discussed.

Fig. 5(c) shows in semilog scale current-voltage characteristics measured on another p-GaN/$VO_2$ junction device at various temperatures. The change in the *I-V* characteristics is more drastic in the insulator-to-metal transition (70 °C – 80 °C) region than in the single phase temperature range. The change in the *I-V* could either be caused by the reduction of the series resistance, barrier height, or conduction mechanism. In the following section, we quantitatively analyze the current-voltage curves of the $VO_2$/GaN junction to identify the possible mechanisms.

Since $VO_2$ has a large carrier density even in the insulating phase and the current is dominated by the current injection from GaN to $VO_2$, thermionic emission mechanism is used to model the hole injection into both the insulating and metallic $VO_2$. If the thermionic emission is the dominant current transport mechanism, the current-voltage curves could be described as[49]

$$I = I_{TE} \exp\left(\frac{q(V - IR_S)}{nkT}\right)\left[1 - \exp\left(-\frac{e(V - IR_S)}{kT}\right)\right] \quad [1]$$

$$I_{TE} = SgA^*T^2 \exp(-q\phi_{b0}/kT) \quad [2]$$

where $R_s$ is the series resistance, $k$ the Boltzmann constant, $q$ the electron charge, $n$ the ideality factor, $I_{TE}$ the saturation current, $S$ the junction area of the diode, $A^*$ the effective Richardson constant (~ 104 A cm$^{-2}$ K$^{-2}$ for $p$-GaN [50]), and $\phi_{b0}$ the apparent barrier height of the junction. The modified voltage, $V' = V - IR_S$, is the voltage drop on the junction, and could be calculated by taking account the external resistance. The above equations directly give

$$\ln\left\{I/\left[1 - \exp\left(-\frac{eV'}{kT}\right)\right]\right\} = \ln I_{TE} + \frac{q(V - IR_S)}{nkT} \quad [3]$$

$$\ln(I_{TE}/T^2) = \ln(SgA^*) - q\phi_{b0}/kT \quad [4]$$

which means that the ideally factor n and $I_{TE}$ can be extrapolated from the slope and intercept of $\ln\left\{I/\left[1 - \exp\left(-\frac{eV'}{kT}\right)\right]\right\}$ versus V curve.[51] Fig. 6(a) shows the semilog plots of $I/\left[1 - \exp\left(-\frac{eV'}{kT}\right)\right]$ as a function of external voltage at 25 °C and 100 °C, which show linear relation in a semilog scale. In this bias range (-0.6 V to 0 V), the current is small enough, so that the contribution to the total voltage from the series resistance could be neglected (the voltage drop on series resistance is less than 1% of the total voltage). The ideality factor at different temperatures can be calculated from the slope of the linear fitting of the curves in Fig. 6(a) and is shown in Fig. 6(b). The ideality factor of the junction increases with rising temperature initially in both the insulating and metallic phase. However, in the temperature range where the insulator-to-metal transition happens, the ideality factor begins to decrease and an obvious hysteresis exists in the heating and cooling curves. One apparent explanation of the decreasing ideality factor can be the decreasing $VO_2$ series resistance across the insulator-to-metal transition, but the current is small enough in the reverse bias region to omit the contribution from the series resistance. In fact, the decrease in ideality factor could be related to the change of the interfacial states caused by the structural transition accompanying metal-insulator transition. The metallic phase $VO_2$ has a rutile structure, while the insulating phase $VO_2$ is in a monoclinic structure with lower symmetry. When going though the metal-to-insulator transition, multiple grains with different orientations can form in the $VO_2$ single crystal and twin boundaries form between these grains. These twin boundaries could induce interfacial states due to any dangling-bond defects. As a result, the metallic phase $VO_2$/GaN interface could have lower interfacial states density and a lower ideality factor than the insulating phase. It has been noted that when $VO_2$ is grown on single crystal substrates with 3*m* surface symmetry (the case of GaN), it will naturally form rotational and twin domains in $VO_2$.[52] In addition, the lattice mismatch between $VO_2$ and GaN is rather large (~ 10 % along the GaN *a* axis).

From intercept of the $I/\left[1-\exp\left(-\frac{eV'}{kT}\right)\right]$ versus $V$ curve, we could also extrapolate $I_{TE}$ at different temperatures, from which the apparent barrier height could be calculated and shown in Fig. 6(c). The magnitude of the barrier height is smaller than but in fair agreement with the Anderson model considering non-ideal factors and systematic error of calculating the barrier height from *I-V* measurements. What is interesting is that the apparent barrier height increases monotonically by 0.17 eV with temperature. Additionally there is only a small hysteresis in the apparent barrier height versus the temperature loop. Within the Anderson model, the change in the barrier height is related to the change in the work function of $VO_2$. Therefore the above observation corresponds to a decrease of the work function of $VO_2$ by ~0.2 eV when going through the insulator-to-metal transition in agreement with previous experiments[10-13], noticing that the Fermi level of GaN is almost temperature independent over this narrow range (the change of the Fermi level of the Mg-doped GaN within this temperature range is much smaller than 0.1 eV[43]). The extrapolated work function of $VO_2$ does not seem to change drastically when going through the phase transition, but evolves rather continuously with temperature. We note to the reader that the idealized Anderson model does not take account into Fermi level pinning and is an inherent limitation.

The change of ideality factor could also be observed in the forward bias region as well. The ideality factor n for forward bias ($V \geq 3k_BT/q$) could be calculated as

$$n(V,T) = \frac{q}{kT}\frac{dV'}{d\ln I} \qquad [5]$$

where ideality factor is a function of temperature and applied voltage. Note that in the forward bias region, the contribution from the series resistance could not be neglected. Therefore we have corrected the series resistance in the calculation. Fig. 7(a) and (b) show how the ideality factor of forward bias region changes with increasing and decreasing temperature, respectively. At a specific temperature, the ideality factor increases with forward bias. In a typical semiconductor p-n junction, an ideality factor of 1 is expected for TE transport mechanism while an ideality factor of 2 is expected for generation-recombination (in the depletion region) mechanism. Typically an ideality factor between 1 and 2 is usually observed and is often explained in terms of a combined transport mechanism in the junction region. For GaN based p-n junctions, however, large ideality factors have been reported in literature and the change in the ideality factor could not be interpreted solely based on the transport mechanism,[53-56] but could be related to other facts such as tunneling current and interface states in these structures.[57] Therefore, the increase of ideality factors with forward bias in Fig. 7 may not be simply explained using current transport mechanism, but could be caused by the interfacial states present in the junction.[57] What is important in Fig. 7, however, is that the ideality factor decreases when $VO_2$ transitions from the insulating state to the higher symmetry metallic state, similar to what is observed in the reverse biased region in Fig. 6(b). Again this could be due to the structural transition in $VO_2$ and may not be related simply to series resistance. In addition, the temperature at which the ideality factor decreases most drastically is ~ 78 ºC and ~ 70 ºC for heating and cooling respectively, close to the transition temperatures determined from the *R-T* measurements. As a result, the change in the ideality factor of the p-n junction could indeed be linked to the phase transition in $VO_2$. One may also note that the ideality factor has slightly different temperature dependence at large forward bias than that at negative biases. In fact, temperature dependence of ideality factor at different biases have been observed before in wafer bonded n-type GaAs and p-type GaN junction.[57]

### C. Frequency dependent capacitance-voltage characteristics of the GaN/VO$_2$ junction

#### a. Reverse-biased region

Fig. 8(a) shows the capacitance-voltage curves of the p-n junctions at various frequencies measured at room temperature. Using the abrupt junction model and depletion approximation, the capacitance of a p-n heterojunction $C_j$ as a function of applied voltage $V$ in the reverse biased region is given by:

$$\frac{S^2}{C_j^2} = \frac{2}{e}\left(\frac{1}{N_{A,GaN}\varepsilon_{GaN}} + \frac{1}{N_{D,VO_2}\varepsilon_{VO_2}}\right)(V_{bi} - V) \quad [6]$$

where $N_{A,GaN}$ is the ionized accepter concentration in GaN, $N_{D,VO_2}$ ionized donor concentration in $VO_2$, $\varepsilon_{GaN}$ and $\varepsilon_{VO_2}$ the dielectric constant of GaN and $VO_2$, $V_{bi}$ the built-in potential and $S$ the electrode area.

For ideal p-n junctions, when reverse biased, the capacitance should be frequency independent up to high frequencies, because the time scale for the space charge layer to respond to external bias is determined by the dielectric relaxation time (majority carrier response time). The dielectric relaxation time ($\tau = \varepsilon / \sigma$, where $\varepsilon$ is absolute permittivity and $\sigma$ the conductivity of the material) of p-type GaN and insulating VO$_2$ is estimated to be ~ 10$^{-12}$ s and ~ 10$^{-11}$ s, respectively, suggesting the capacitance of the junction should be frequency independent to gigahertz frequencies. The small-signal capacitance under reverse external biases in Fig. 8(a), however, exhibits strong frequency dependence even at relatively low frequencies. Non-idealities such as traps, series resistance and minority carriers can significantly affect the measured small signal capacitance and conductance.[58-64] In the present case, the influence of bulk traps could be eliminated because neither the capacitance of VO$_2$ thin films nor GaN substrates shows any strong frequency dependence within the range studied (see figure 8(c) and (d) for reference). On the other hand, factors such as interface states, series resistance and minority carriers would contribute as suggested from previous current-voltage measurements. Usually, a small signal equivalent circuit model is used to consider the interface and series resistance effect,[64] and the resonance frequency is determined by the parameters in the circuit model. In the reverse bias region, the effective circuit parameters are related to the series resistance and interfacial states, whereas the minority carrier dynamics will contribute to the forward bias region as well. As a result, comparing the resonance frequencies at different bias regions would provide information on the minority carrier dynamics.

### b. Forward-biased region

The capacitance of a forward biased ideal p-n junction is dominated by the diffusion capacitance and could be highly frequency-dependent because of the minority lifetime. Fig. 9(a) compares the frequency dependence of the measured capacitance under forward and reverse bias at 25 ºC. A clear difference in the frequency range where the capacitance begins to drop could be identified. Because the frequency dependence of capacitance under reverse bias is primarily due to interface states and series resistance, while the forward bias capacitance is a result of all the contributions from the interface states, series resistance, and minority carriers, we speculate that the shift in the time constant should be caused by the slow minority dynamics in vanadium oxide. Recently it has been shown that the minority carrier diffusion length is on the order of a few μm in $VO_2$ using scanning photocurrent microscopy, indicating that the minority carrier lifetime is roughly 4–15 microseconds.[65] It was conjectured that such long minority carrier lifetime is because the conduction and valence band of $VO_2$ are both derived from the d bands as shown in Fig. 2(b) and therefore the electron-hole recombination through the first-order dipole transition is forbidden.[65] The resonance frequency in Fig. 9(a) corresponds to a time constant of $\tau_0 = 1/f_0 = 6\mu s$, in the range of minority carrier lifetime determined by the previous study, suggesting that the frequency dependence of forward capacitance is indeed due to the minority carrier dynamics. In contrast, Fig. 9(b) shows the frequency dependence of measured capacitance under different dc bias. There is no obvious difference in the *C-f* characteristics between the forward and reverse bias region. The above observation indicates that the minority carrier dynamics in metallic $VO_2$ is much faster than that in the insulating phase. This can be caused by the several orders of magnitude increase in the carrier density. The frequency dependence in the metallic phase should mainly be originated from the series resistance and interfacial states. Therefore the above observations support that the minority carrier dynamics is mainly responsible to the frequency shift in the insulating phase. Below we quantitatively analyze the minority carrier contributions. Note that the total capacitance measured in the metallic phase is larger than that measured in the insulating phase, which is related to the increase in the dielectric constant of vanadium dioxide (the capacitance of $VO_2$ is in series with the junction capacitance).[45]

When a p-n junction is biased by a forward dc voltage $V_0$ plus an ac signal $V_1 e^{i\omega t}$, the current density flowing through the junction can be expressed as

$$J(t) = J_0 + J_1 e^{i\omega t} \qquad [7]$$

where $J_0$ is the dc current under $V_0$, and $J_1$ is the ac current component, which can be a complex number reflecting the phase lag between current and voltage. The conductance $G$ and diffusion capacitance $C$ is given by

$$Y \equiv \frac{J_1}{V_1} \equiv G + i\omega C \qquad [8]$$

At high frequencies, the recombination of minority carriers could not follow the external voltage and the minority carrier current will go out of phase with respect to the applied voltage, leading to strong frequency-dependence of the capacitance when the ac signal frequency approaches the minority lifetime.

In the present study, because of the high injection efficiency in $VO_2$/GaN junction, we can simplify the system as a one-sided junction considering only the holes injected into $VO_2$. By assuming that the quasi-Fermi level of the holes is a constant through the p-type GaN side to the whole space charge layer of the junction, i.e. there is no minority carrier recombination in the space charge region, we can only consider the carrier dynamics in the neutral region outside of the space charge region of $VO_2$. The equation of state describing hole concentration on the $VO_2$ side is:

$$\frac{\partial \Delta p}{\partial t} = D_p \frac{\partial^2 \Delta p}{\partial x^2} - \frac{\Delta p}{\tau_p} \quad [9]$$

where $\Delta p$ is the difference between hole concentration at location $x$ at time $t$ and the equilibrium hole concentration ($\Delta p = p(x,t) - p_0$), $D_p$ the diffusion coefficient of holes in $VO_2$, $\tau_p$ the minority carrier lifetime in $VO_2$. The boundary condition is set by the applied DC bias $V_0$ and AC bias $v(t) = V_1 e^{i\omega t}$ by $\Delta p(x=0,t) = p_0[\exp\left(\frac{q(V_0 + V_1 e^{i\omega t})}{kT}\right) - 1]$, where x =0 corresponds to the edge of the space charge region of $VO_2$. In the case of $|V_1| = kT$, the above boundary condition could be simplified as $\Delta p(x=0,t) = p_0[\exp\left(\frac{qV_0}{kT}\right)\left(1 + \frac{V_1 e^{i\omega t}}{kT}\right) - 1]$. Another boundary condition typically used is $\Delta p(x=\infty,t) = 0$. In addition, we assume that the initial condition is the steady state condition under the DC voltage $V_0$, which is given by $\Delta p(x,t=0) = p_0[\exp\left(\frac{qV_0}{kT}\right) - 1]\exp\left(-\frac{x}{L_p}\right)$, where $L_p = \sqrt{D_p \tau_p}$ is the diffusion length of holes in $VO_2$.

The classical way to simplify the equation is to assume that $\Delta p(x,t) = \Delta p(x)e^{i\omega t}$ and the ac current will be

$$J_1 = \frac{qD_p p_0}{L_p}\sqrt{1+i\omega\tau}\left[\exp\left(\frac{qV_0}{kT}\right)\right]\frac{qV_1}{kT} \quad [10]$$

For low frequencies $\omega\tau = 1$, the conductance and capacitance corresponds to the dc values. In the limit of high frequencies $\omega\tau ? 1$, the diffusion capacitance is proportional to $(\omega\tau)^{-1/2}$.

As shown by the dashed line in Fig. 9(a), The relation $C \propto (\omega\tau)^{-1/2}$ captures the frequency dependence of the capacitance in the intermediate frequency region, but does not explain the drop of capacitance at higher frequencies. On the other hand, if an effective small signal circuit model is applied,[64, 66] the drop of the capacitance near the resonance frequency can be predicted.

The reason why classical theory may not be accurate is that the net charge in the diffusion region may not be zero, leading to conduction modulation. Because of the long minority carrier lifetime, the holes can accumulate in $VO_2$ and the low injection level condition is not valid any more (when $\Delta p \mu_p$ is on the same order of magnitude of $n_0 \mu_n$ in $VO_2$). This will modulate the electric field distribution, so the assumption that $\Delta p$ follows the external bias by $\Delta p(x,t) = \Delta p(x)e^{i\omega t}$ independent of location is not strictly valid and the equation of state needs also to be modified. Therefore charge neutrality may not be assumed in the diffusion region and the drift component needs also be considered.[67] This would lead to a phase lag between the current and voltage and impedance behavior of the junction.[65, 69] In addition, since the thickness of $VO_2$ layer (25 - 40 nm) is much smaller than the minority carrier diffusion length (a few micrometers), the boundary condition used in the classical condition also needs to be corrected. What's more, the quasi-Fermi level of the holes may not be flat across the space charge region if the minority carriers in $VO_2$ can follow external voltage. Since the existing models have separately considered effects such as conductance modulation and short device length, a more generic model take into account of all these effects needs to be developed. Understanding the minority carrier dynamics and its consequence in correlated oxides could be of general interest for the study on whether and how the electron correlation could be changed by the excess minority carrier injection. Despite the limitations of such models, the carrier lifetime must be of the same order of the resonance frequency at forward bias for such a frequency shift to happen. On the other hand, Fig. 9(b) suggests that the hole carrier lifetime is less than 1 μs in the metallic phase due to the larger electron density.

Interestingly, we notice that the overall measured 'capacitance' becomes negative at high frequencies under and only under forward bias (note that the 'negative' capacitance simply means the measured current lags behind the applied voltage by a phase of 0 to π/2 as can been seen from equation [8]). The negative capacitance phenomena have in fact been observed in different kinds of p-n junctions[66-78], and there are various explanations including minority carrier lifetime[64, 66, 77], the inaccuracy of the classical I-V formula of pn junction[67] and extrinsic artifacts such as series resistance and parasitic impedance[61, 68]. Many of the negative capacitance phenomena, for example, have been observed in low frequency ranges (few Hz to kHz range) rather than high frequencies, which seems not be the case for the present study.[70, 71, 76, 78] In addition, the series resistance effect and interfacial states may also not be a major factor in the present work because there is no such behavior at high temperatures as shown in Fig. 9(b). Hence the observed negative capacitance may not be related to extrinsic artifacts, because these factors should not change significantly with the device temperature over the range studied here. This suggests that the negative capacitance could be intrinsic to the device and be related to the minority carrier dynamics in $VO_2$. Under high frequencies, the recombination of minority carriers could not follow the external voltage and the minority carrier current will go out of phase to the applied voltage due to minority carrier storage and conductance modulation.[64, 66, 77] The apparent capacitance becomes negative because of the phase lag between current and voltage. The nonlinearity of the *I-V* curves of the bulk insulating $VO_2$ may also contribute to such effect.[78] Fig. 9(c) shows the contour plot of the capacitance-voltage-frequency curve measured at 25 ºC. A clear shift in the frequency could be observed once the junction is forward biased, suggesting the role of minority carriers.

### c. Built-in potential from C-V measurements

In principle, the built-in potential can be extrapolated from the intercept of the $1/C^2$-*V* curves. However, the frequency-dependence of capacitance will make the calculation of the barrier height less accurate. Therefore, we could only estimate the built-in potential assuming that the capacitance (at 1k Hz) begins to drop when the external bias cancels the built-in potential as shown in Fig. 8(a) (which is also the voltage where the conductance increases dramatically). The barrier height is estimated to change from 1.4 eV to 1.6 eV when the temperature changes from 25 ºC to 100 ºC. This corresponds to a decrease of ~ 0.2 eV in the work function of $VO_2$. In comparison to the results in Fig. 7(a) from current-voltage measurements, the absolute value of barrier height from *C-V* is larger because any parallel conduction mechanism can lead to underestimation of the barrier height in *I-V* measurements. The evolution of barrier height as a function of temperature from both measurement techniques is however consistent.

### D.  High bias current-voltage characteristics

We have also studied the *I-V* characteristics of the junction under high bias. It is found that the metal-insulator transition in $VO_2$ can be triggered by voltage at large enough bias (~ 100 V) in the junction. The voltage induced MIT lead to an increase in the junction current magnitude by a few times.[79] The junction *I-V* curves become hysteretic due to such voltage induced metal-insulator transition. In addition, the shape of *I-V* curves are dependent on the sweeping speed of the external voltage. With slower sweeping speed, the increase in the current becomes larger and the *I-V* curves become hysteretic.

### IV    CONCLUSIONS

We have fabricated GaN / $VO_2$ p-n heterojunctions and studied their temperature depedent electrical properties, providing information on the barrier height and minority carrier life time of $VO_2$. We find that the idealiy factor of these junctions changes across the $VO_2$ phase transtion and attribute this to the structrual transtion between two crystal structures with different symmetries, with larger ideality factor for low symmetry phase. The measured evolution of the apparent barrier height and work funtion of $VO_2$ is similar in both current-voltage and capacitance voltage methods across the insulator to metal transtion. The band offset could be qualitatively explained by a classical band bending model with electron affinity rule. By analyzing distinct capacitance-frequency behaviour in the reverse and forward bias regime in two different phases, we separate the miniority carrier effect from the interface states and series resistance effect and estimate the minority carrier lifetime in $VO_2$ to be of the order of few microseconds. The integration of high speed switchable correlated oxides with conventional semiconductor heterojunctions create a new platform to study physics of correlated electron systems, and have implications for future devices based on tunable phase transitions.

**Acknowledgements**

We gratefully acknowledge ONR N00014-12-1-0451 and NSF DMR-0952794 for financial support.

**FIGURE CAPTIONS**

Fig. 1 (a) The electronic band structure of n-type phase change insulator and p-type semiconductor before the two materials are brought together to form a p-n junction. (b) The band structure of the insulator can be modified by inducing phase transitions driven by thermal, strain, electrical or optical perturbations. The collapsed band gap can lead to different properties of the junction. (c) and (d) The depletion region in the material, $x_n$, can be quite different in the insulating phase than the metallic phase due to the different free carrier density in the two phases. (e) and (f) By controlling the external voltage, one can modulate depletion width and the minority carrier density in the phase change insulator. Since electron correlation is highly dependent on the carrier density, the active control of the voltage can induce different phases at the junction.

Fig.2 The electronic band structure of the (a) metallic phase and (b) the insulating phase of $VO_2$. The parameters such as band gap, band width and work function are taken from experimental and first principles studies. [Ref. 24 - 31, 34 - 36]

Fig. 3 The band diagram of the $VO_2$/GaN n-p junctions as predicted by Anderson model for (a) metallic and (b) insulating $VO_2$. Assuming depletion approximation model, GaN forms a Schottky junction with metallic $VO_2$ and a type I p-N junction with insulating $VO_2$. The Schottky barrier height and conduction/valence band offsets are shown in (a)

and (b), respectively. The shaded part corresponds to energy bands, and the blank part corresponds to energy gap.

Fig. 4 (a) Resistance versus temperature curve measured in-plane from the VO2 grown on p-type GaN, with the inset showing measurement geometry. The leakage current through the p-GaN substrate is negligible. (b) The d$\ln R$/d$T$ versus temperature curves for heating and cooling branches. The transition temperature is estimated to be 75 ºC and 68 ºC for the heating and cooling, respectively from the Gaussian fit of the peaks.

Fig. 5 (a) The measured current-voltage curves of the GaN/VO$_2$ junctions at 25 ºC and 100 ºC. Rectifying behavior can be observed at all temperatures. The device is biased in such a way that VO$_2$ is grounded and voltage drops on GaN as drawn in the inset. (b) The current-voltage characteristics of the Pd/GaN/Pd contacts. The linearity of these curves at different temperatures indicates the formation of ohmic contact between Pd and GaN. (c) The current-voltage characteristics of the GaN/VO$_2$ heterojunction at various temperatures plotted in a semilog scale. The large reverse bias current could be related to the tunneling through the interfacial states.

Fig. 6 (a) Semilog plot of the $I/\left[1-\exp\left(-\frac{eV'}{kT}\right)\right]$ versus $V$ curves for reverse bias region at 25 °C and 100 °C. (b) Ideality factor as a function of temperature calculated from the slope of the $\ln\left(I/\left[1-\exp\left(-\frac{eV'}{kT}\right)\right]\right)$ versus $V$ curves. A clear drop in the ideality factor and hysteresis could be observed. The hysteresis width agrees with that from the resistance-temperature measurements of $VO_2$. (c) The apparent barrier height calculated from the $I$-$V$ characteristics at different temperatures.

Fig. 7 The ideality factor as a function of temperature extrapolated from the forward bias region during (a) heating and (b) cooling. A clear decrease in the ideality factor can be observed across the phase transition temperature in each branch.

Fig. 8 (a) The capacitance-voltage curves of the p-n junction under different frequencies at 25 °C. The frequency dependence of the reverse-biased capacitance is an indication of the interfacial states and series resistance in the system. (b) The capacitance-voltage curves near the turn-on voltage under different frequencies at 25 °C. At high frequencies, the measured capacitance becomes 'negative' under forward bias. (c), (d) Measured capacitance-frequency dependence of $VO_2$ and GaN thin films, respectively. Neither of them shows strong frequency dependence, suggesting that the polarization due to the bulk defects is not responsible for the frequency dependence in (a).

Fig. 9 (a) The frequency-dependent junction capacitance under various dc bias at 25 ºC. The frequency dependence at reverse bias is related to interfacial states and series resistance. At forward bias, the minority carrier dynamics will also contribute to the measured capacitance, leading to the shift in the resonance frequency of the junction. (b) The frequency-dependent junction capacitance under various dc bias at 100 ºC. In contrast to (a), the measured capacitance exhibits similar frequency dependence in both the reverse and forward biased region, indicating that the minority carrier lifetime in the metallic phase is much smaller than that in the insulating phase due to enhanced electron density.(c) Contour plot of the capacitance-voltage-frequency curve measured at 25 ºC. A clear shift in the frequency could be observed once the junction is forward biased.

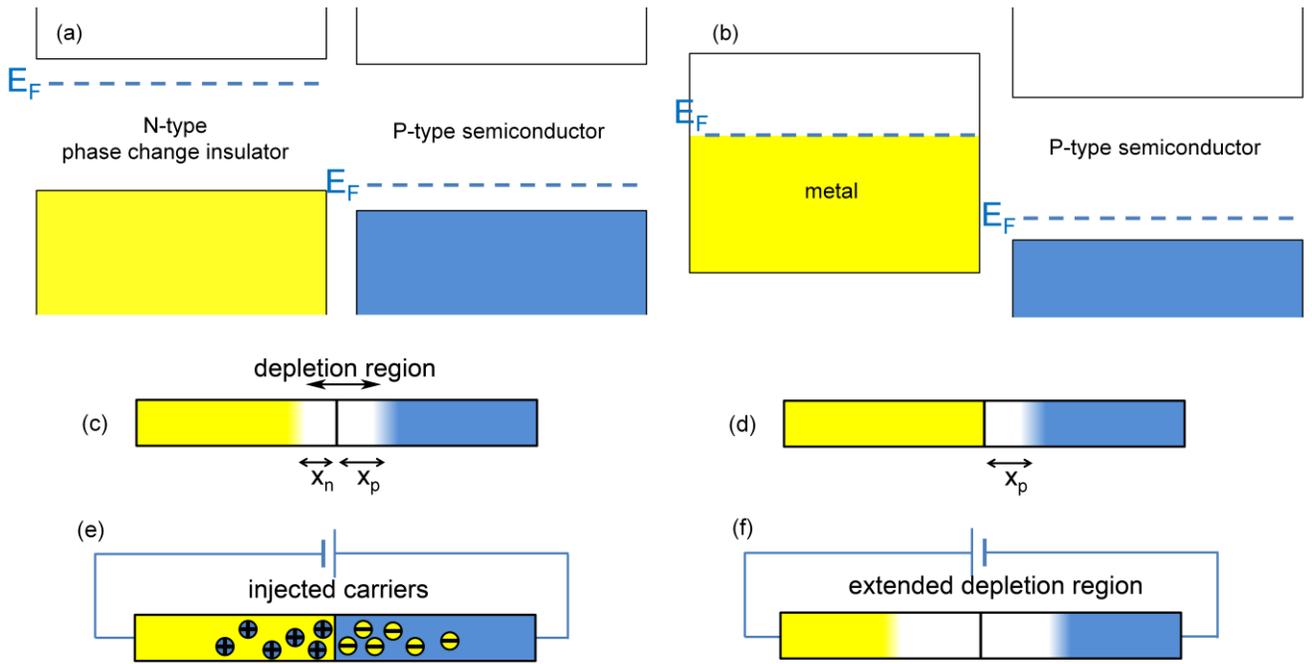

Fig. 1

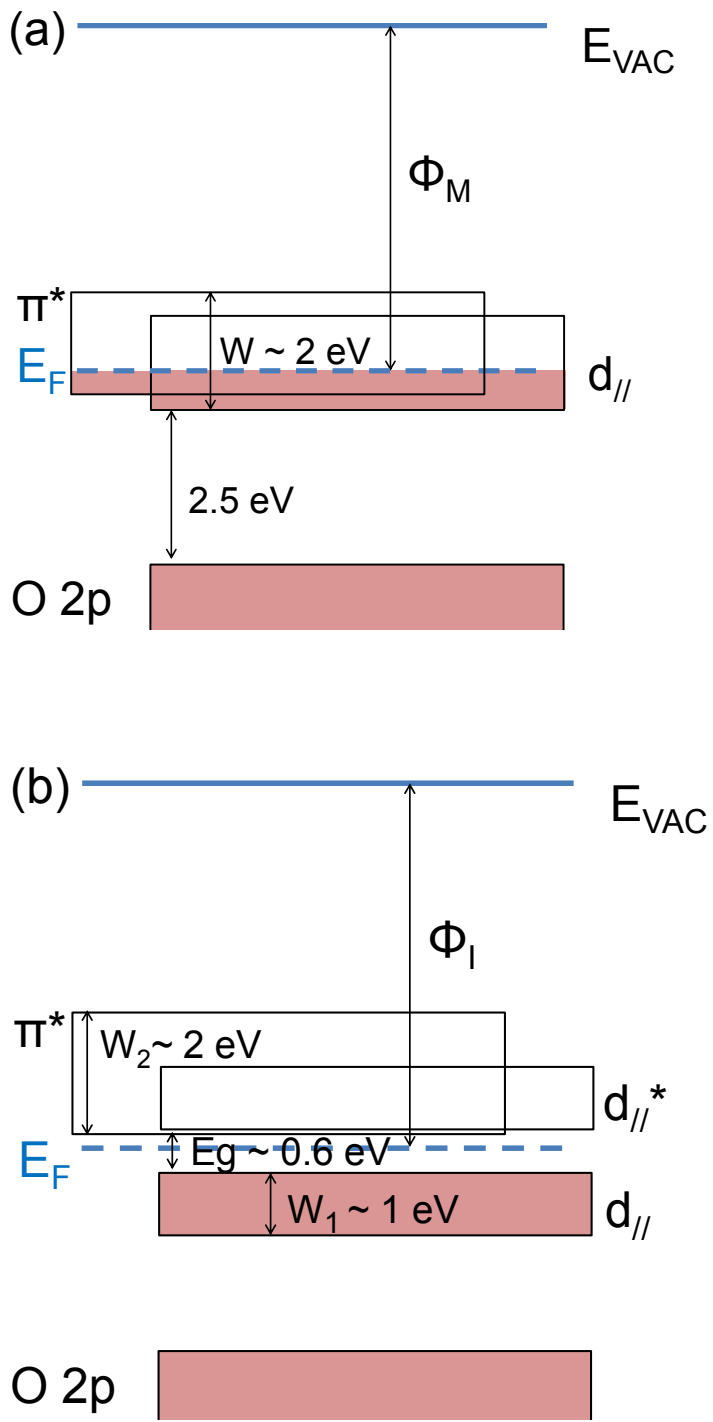

Fig. 2

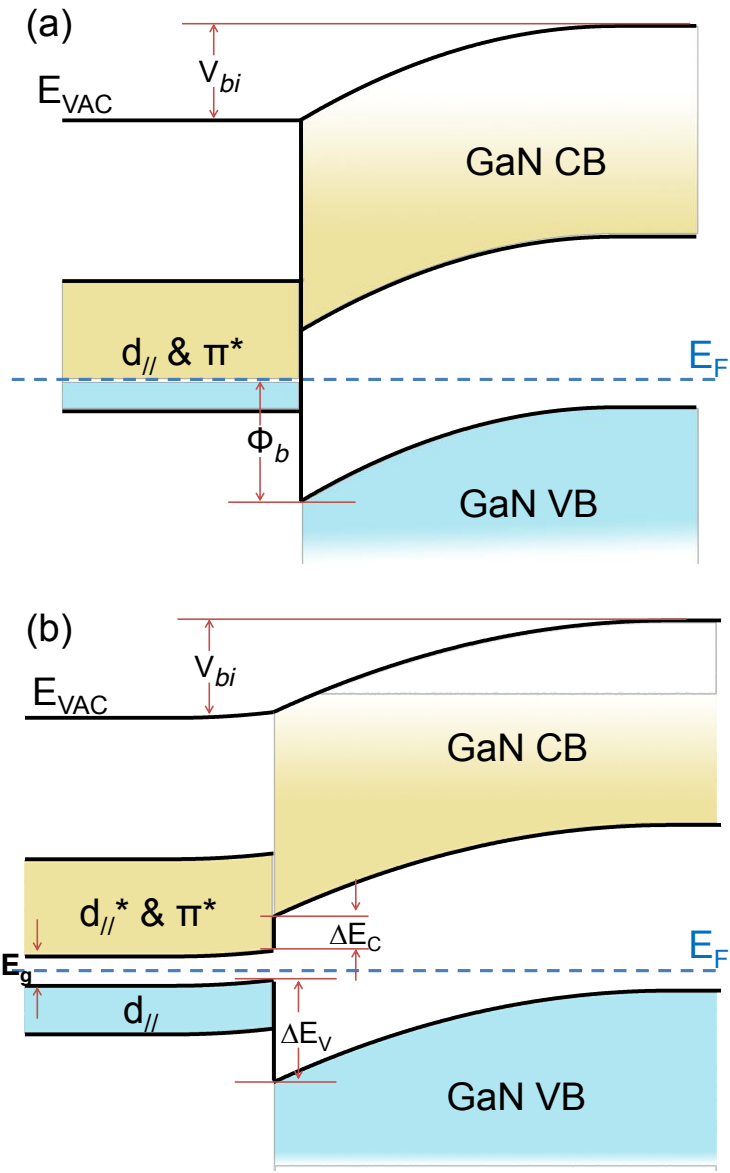

Fig. 3

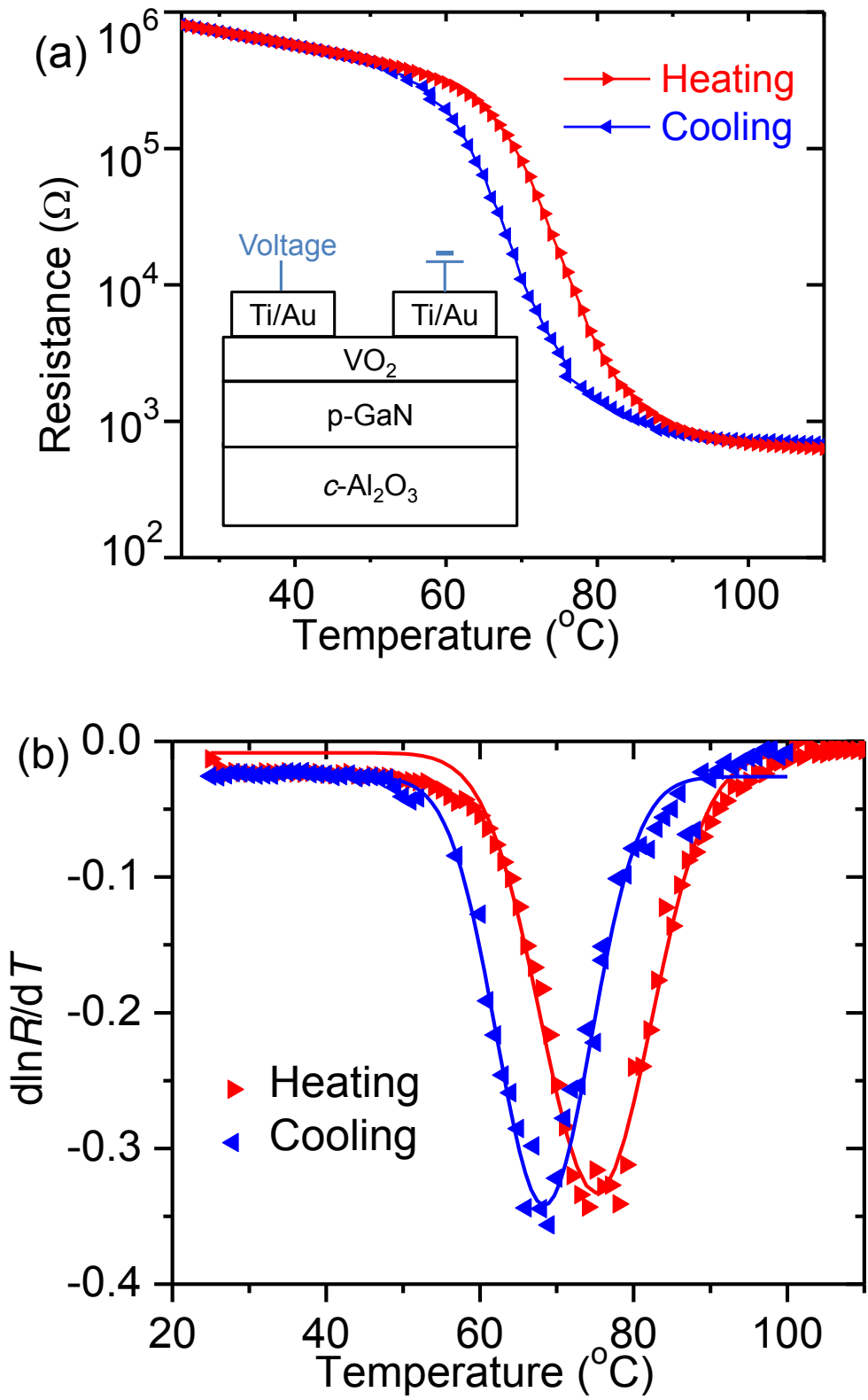

Fig. 4

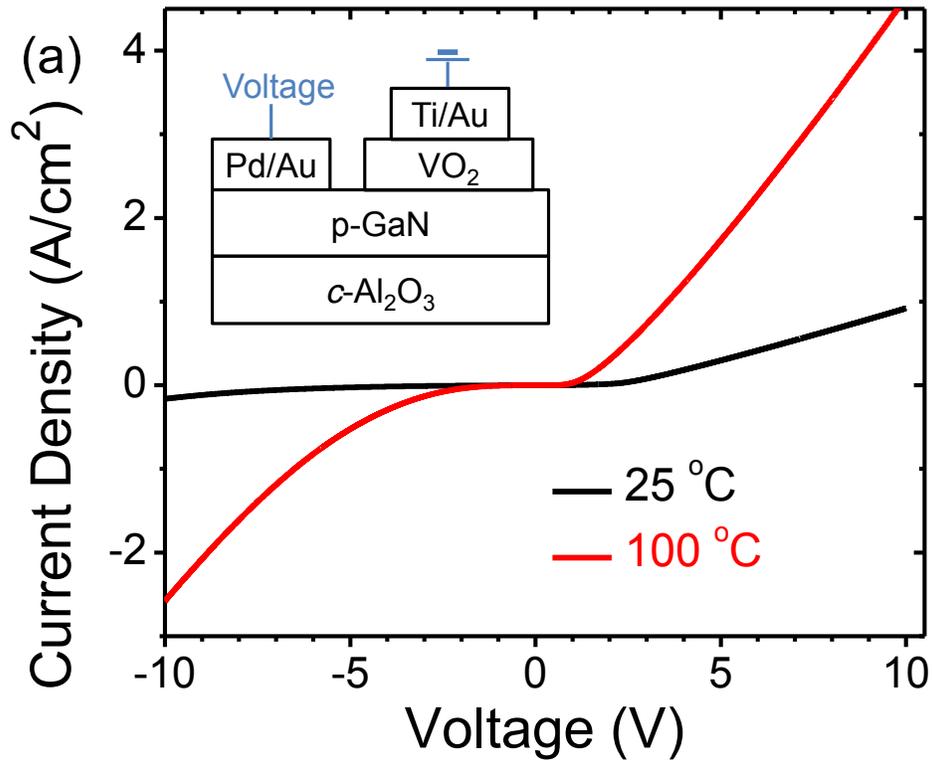

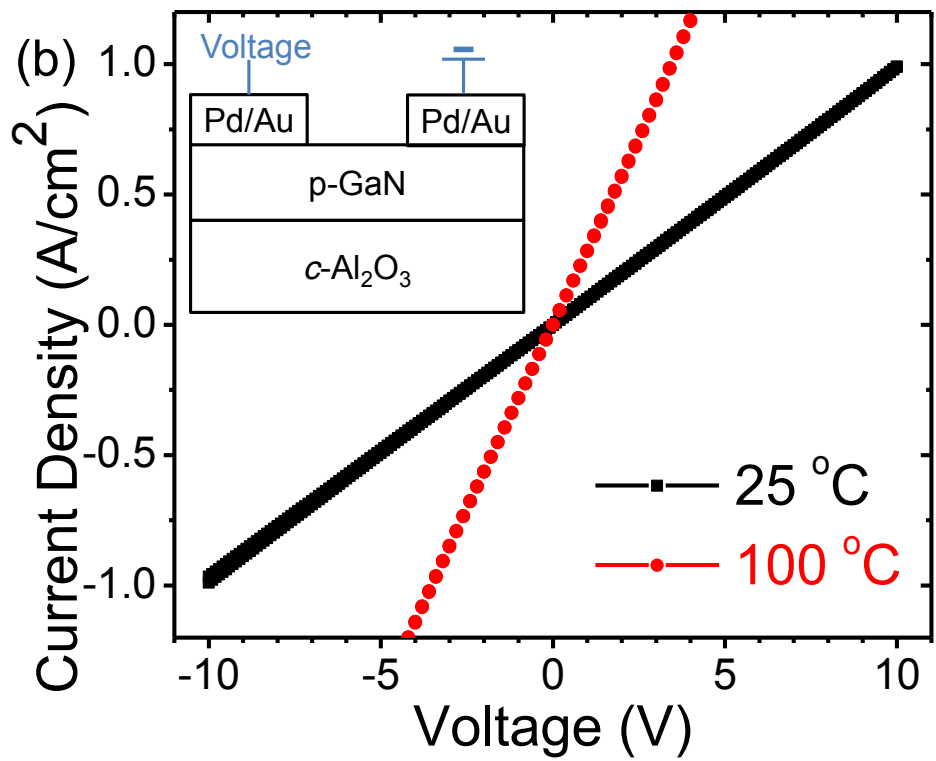

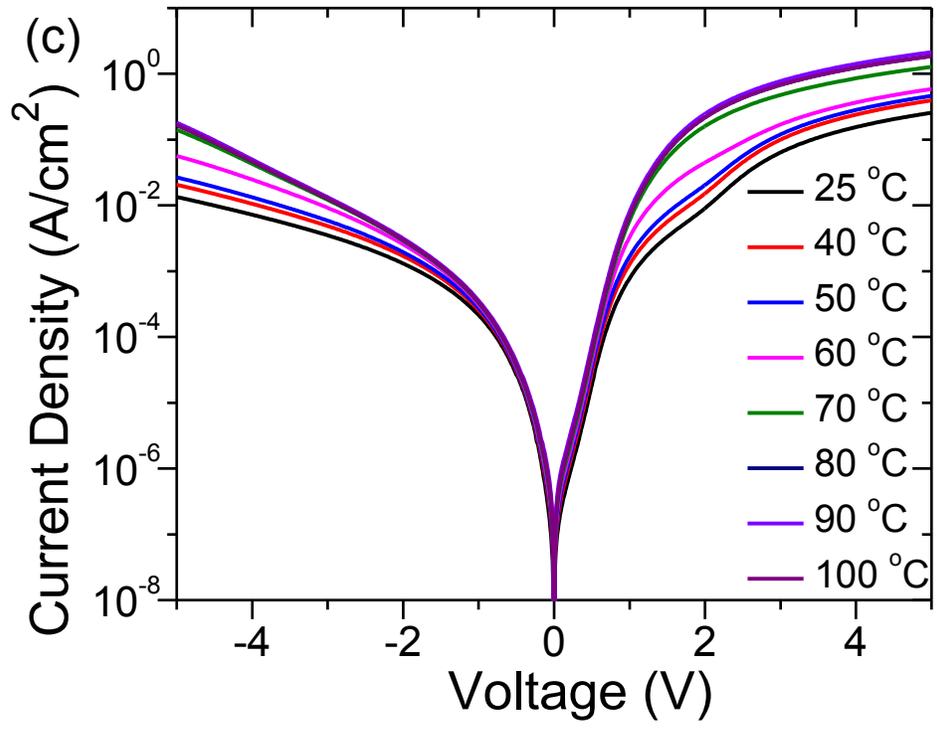

Fig. 5

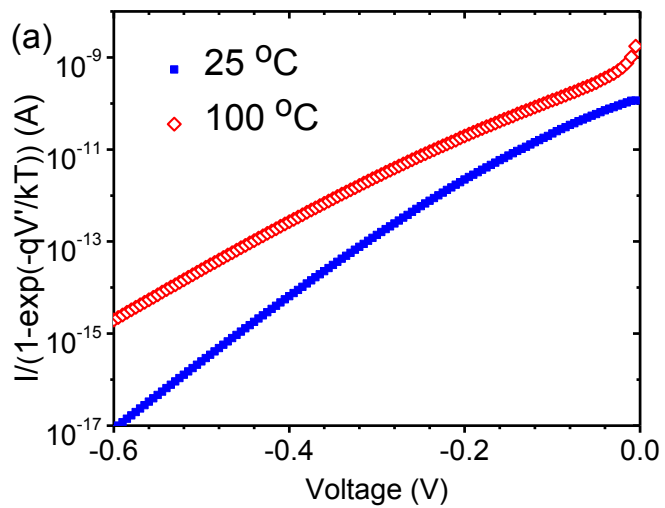
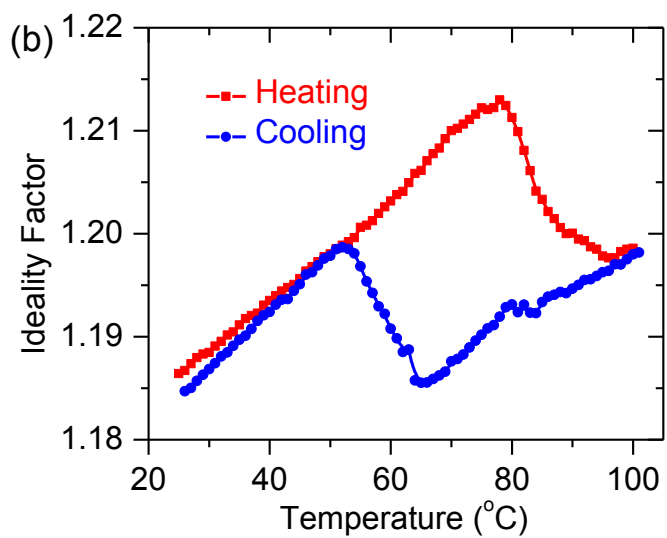
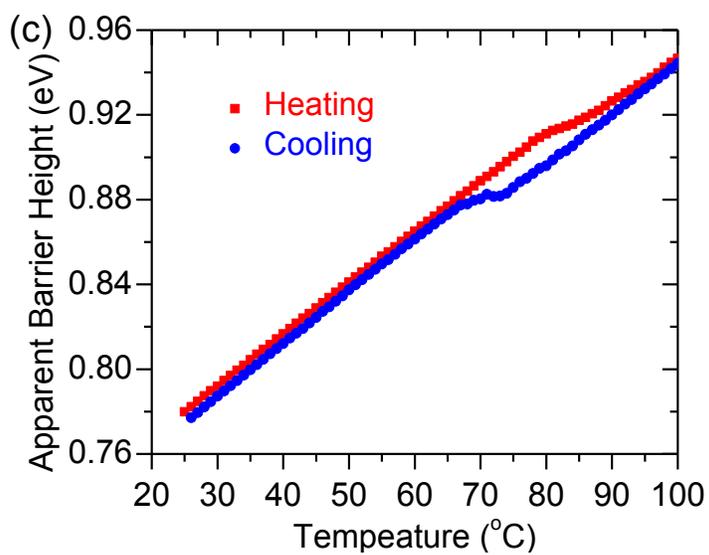

Fig. 6

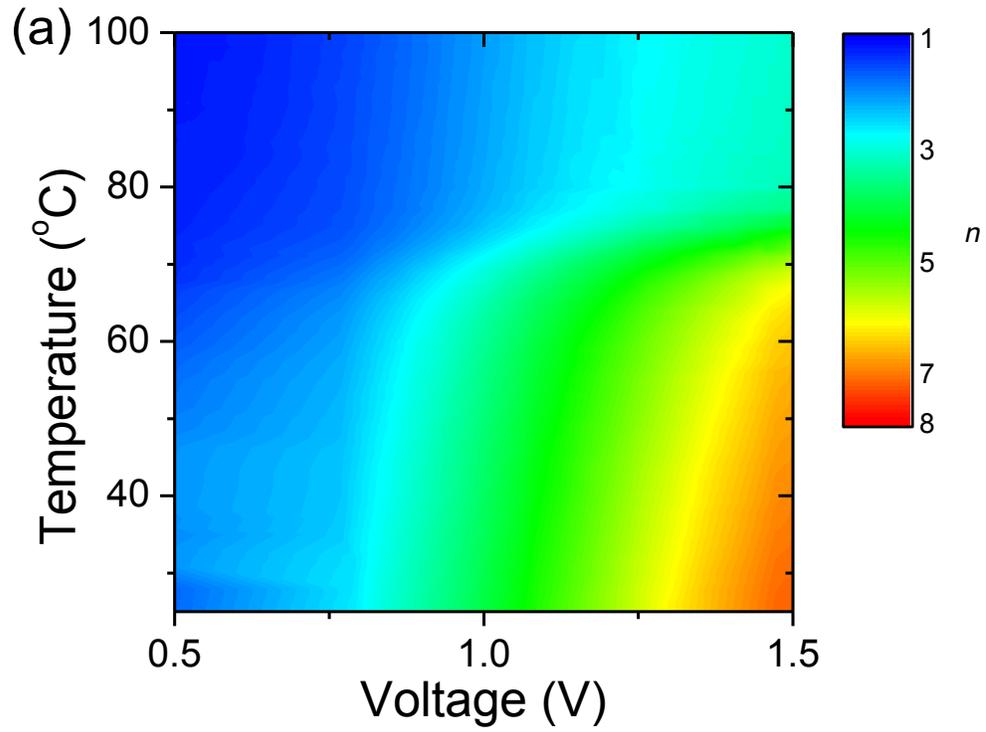

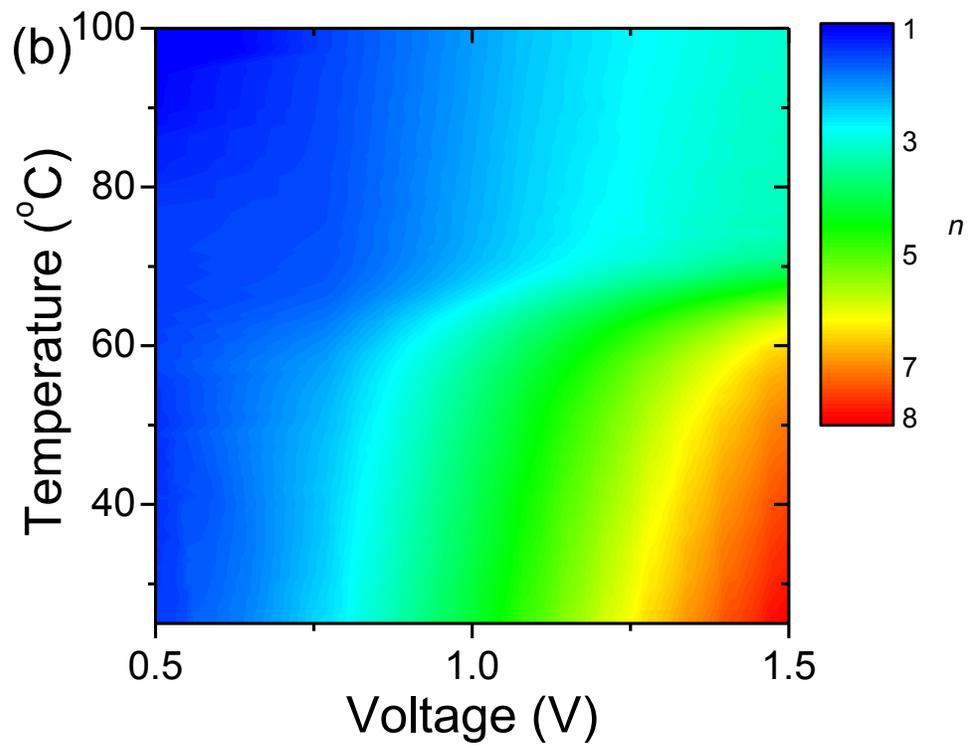

Fig. 7

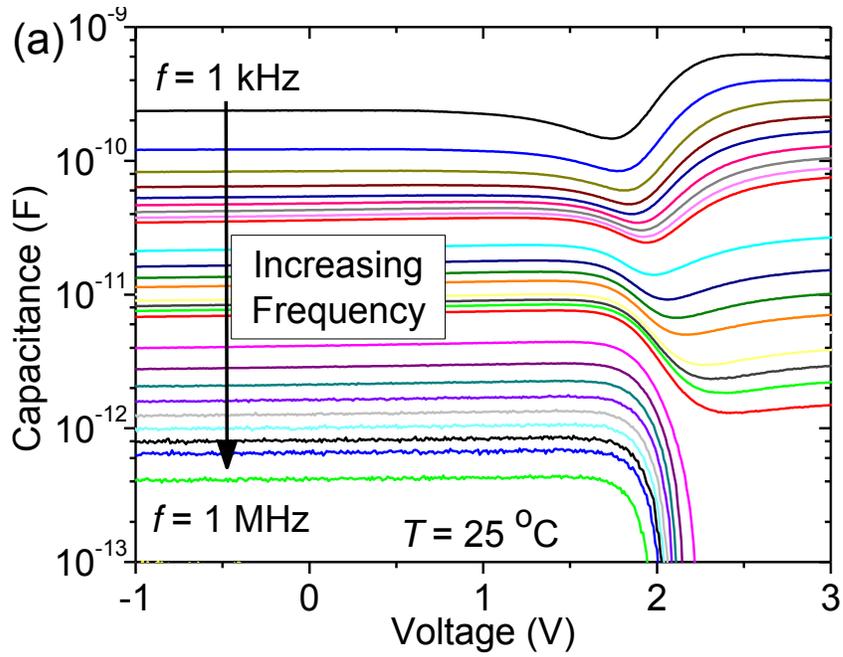

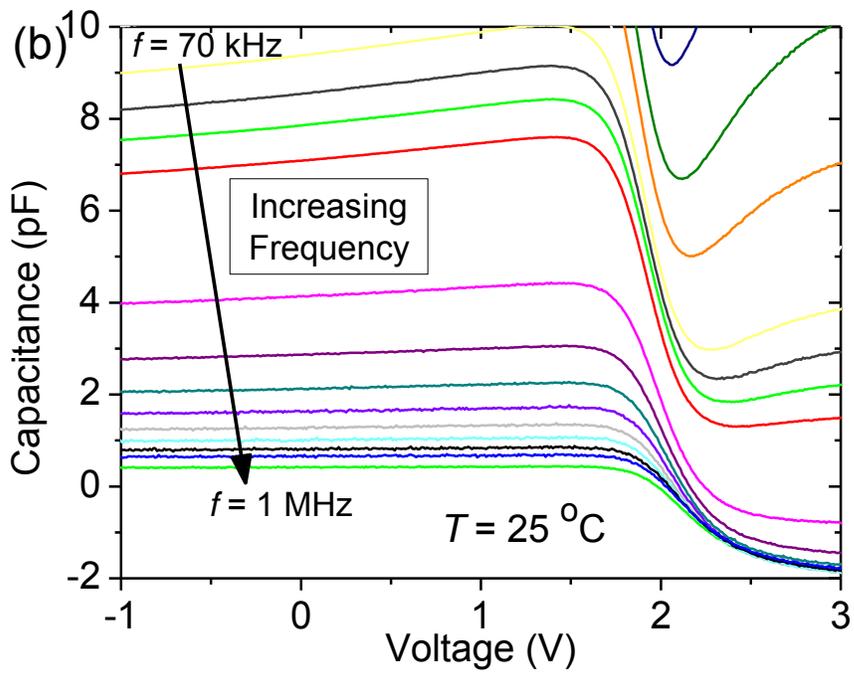

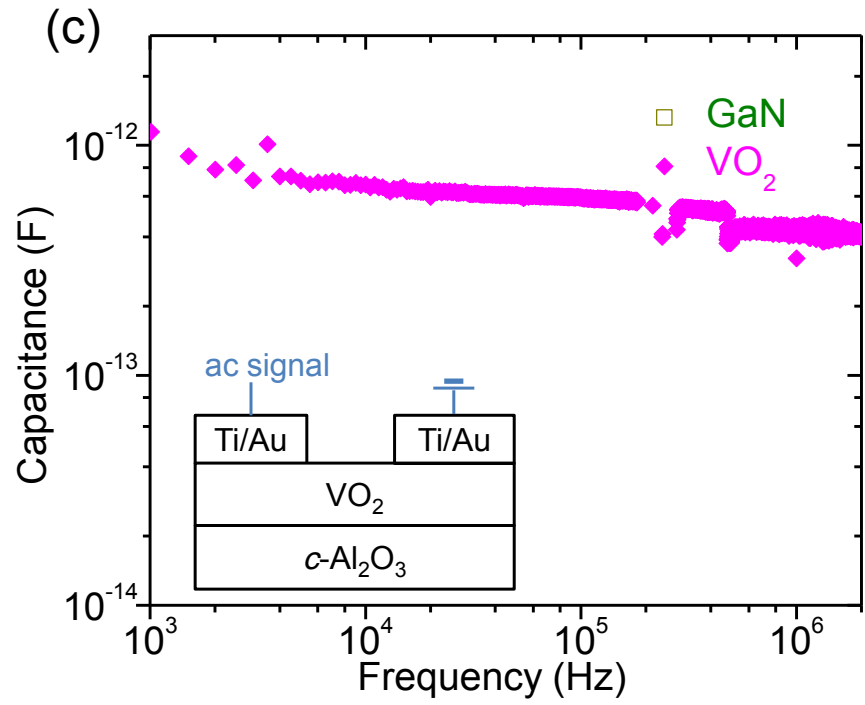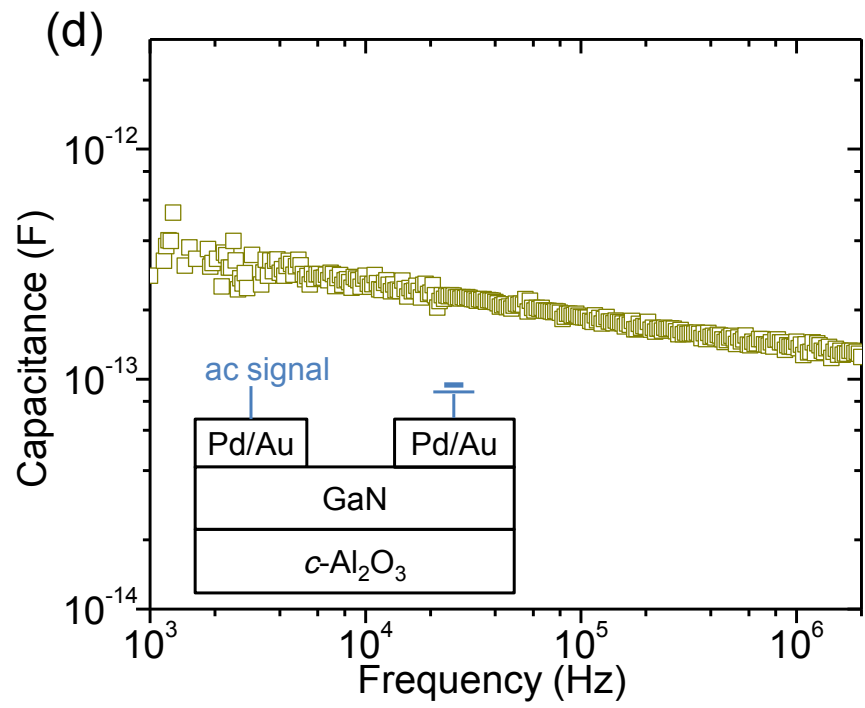

Fig. 8

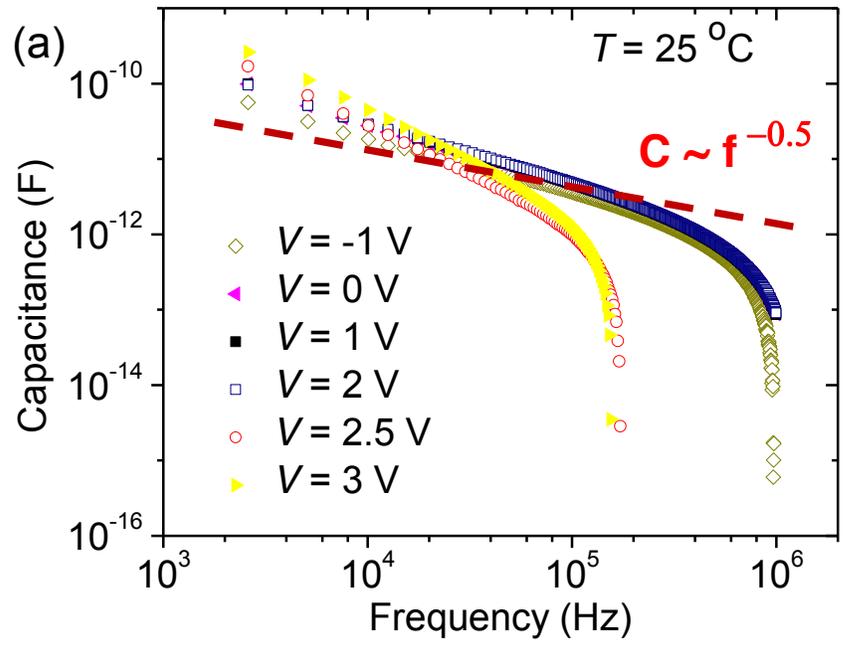

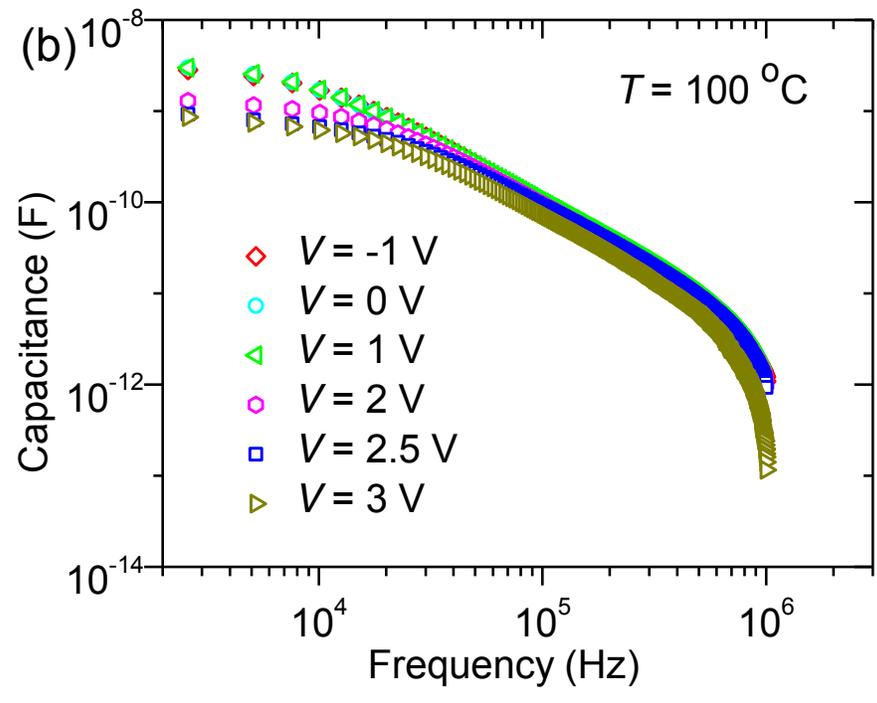

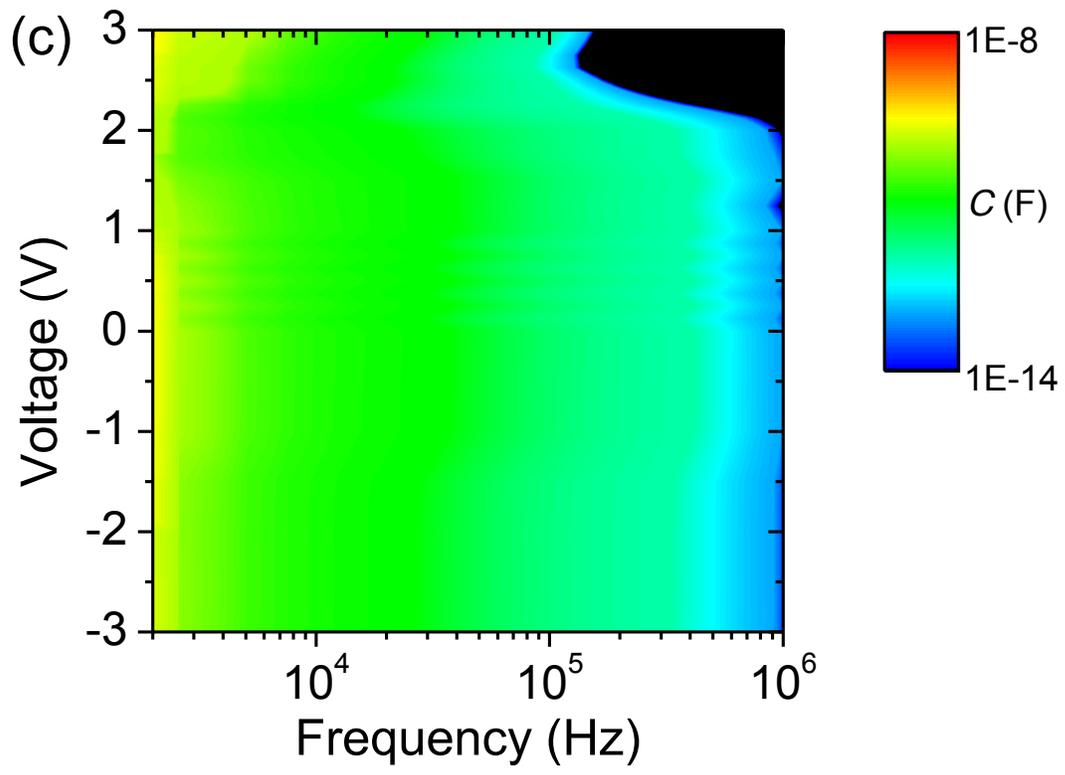

Fig. 9